# Using machine learning to downscale coarse-resolution environmental variables for understanding the spatial frequency of convective storms [*]

Hungjui Yu [a,b], Lander Ver Hoef [b], Kristen L. Rasmussen [a], Imme Ebert-Uphoff [b]

[a] *Department of Atmospheric Science, Colorado State University, Fort Collins, CO*

[b] *Cooperative Institute for Research in the Atmosphere, Colorado State University, Fort Collins, CO*

*Corresponding author*: Hungjui Yu, hungjui@rams.colostate.edu

ABSTRACT

Global climate models (GCMs), typically run at ~100-km resolution, capture large-scale environmental conditions but cannot resolve convection and cloud processes at kilometer scales. Convection-permitting models offer higher-resolution simulations that explicitly simulate convection but are computationally expensive and impractical for large ensemble runs.

This study explores machine learning (ML) as a bridge between these approaches. We train simple, pixel-based neural networks to predict convective storm frequency from environmental variables produced by a regional convection-permitting model. The ML models achieve promising results, with structural similarity index measure (SSIM) values exceeding 0.8, capturing the diurnal cycle and orographic convection without explicit temporal or spatial coordinates as input. Model performance declines when fewer input features are used or specific regions are excluded, underscoring the role of diverse physical mechanisms in convective activity.

These findings highlight ML's potential as a computationally efficient tool for representing convection and as a means of scientific discovery, offering insights into convective processes. Unlike convolutional neural networks, which depend on spatial structure and grid size, the pixel-based model treats each grid point independently, enabling value-to-value prediction without spatial context. This design enhances adaptability to resolution changes and supports generalization to unseen environmental regimes, making it particularly suited for linking environmental conditions to convective features and for application across diverse model grids or climate scenarios.

SIGNIFICANCE STATEMENT

Understanding how clouds and storms form is critical for predicting extreme weather in a changing climate. However, current climate models struggle to capture small-scale cloud features due to their coarse resolution. Our research shows that machine learning, trained on detailed weather simulations, can successfully predict where and how often intense storms form based on larger-scale environmental patterns. This approach bridges the gap between what climate models can resolve and the finer details of clouds that matter most for impacts like floods or severe weather. By combining high-resolution simulations and machine learning, our work offers a powerful new way to better understand storms in both current and future climate.

## 1. Introduction

Clouds play a crucial role in the Earth system. From a weather perspective, clouds are central to precipitation and storm formation, making them a primary driver of extreme weather globally (Houze 2018). In the broader climate system, clouds are essential to the global circulation and water cycle, significantly influencing Earth's water and energy budgets (Allen and Ingram 2002; Stubenrauch et al. 2013). The frequency of storm-related weather and climate disasters with over $1 billion USD in economic impacts has significantly risen in the U.S. from 1980 to 2024 (NOAA National Centers for Environmental Information U.S. Billion-Dollar Weather and Climate Disasters 2025). Increasing extreme weather events have been seen globally as well (Stott 2016; Myhre et al. 2019; Li et al. 2019). These trends underscore the urgent need, for both the scientific community and the public, to better understand how cloud systems form, evolve, and interact with the large-scale environment, particularly in a changing climate.

Traditionally, exploring the relationship between clouds and the large-scale environment under changing climates relied on global climate models (GCMs), typically operating at ~100-km grid spacing. However, clouds remain one of the largest uncertainties in climate projections (Change, I. P. O. C., 2007). While GCMs can adequately represent the large-scale environments (Chen et al., 2020), their coarse resolution leads to substantial biases in cloud radiative feedback (Le Treut & McAvaney 2000; Bony et al. 2004) and poor representation of key characteristics such as the seasonal and diurnal cycles of convection, precipitation, and organized cloud systems (Allen and Ingram 2002). These models fundamentally lack the resolution needed to capture the mesoscale and convective-scale structures, and dynamics of cloud systems; thus, they rely on convection parameterizations to represent these processes.

Recent advances in computational resources have enabled convection-permitting regional climate models at horizontal resolutions ~4 km (Kendon et al., 2016). These models explicitly resolve mesoscale processes and have shown substantial improvements in simulating regional storm-scale variability (Prein et al., 2015) and the precipitation diurnal cycle (Rasmussen et al. 2017). Based on previous work, we utilized algorithms to classify convective and stratiform clouds and identify satellite-based Tropical Rainfall Measuring Mission (TRMM)-heritage storm modes within these high-resolution model outputs (Steiner et al. 1995; Houze et al. 2015). This effort has produced extensive cloud type and storm mode databases from long-term convection-permitting simulations over the continental U.S. (13-year, 21-year, and 40-year current and future climate simulations over the continental U.S.; Rasmussen et al. 2023; Liu et al. 2016). These resources have been instrumental in assessing how regional climate models capture convection and how these storm modes might evolve under future climates. Furthermore, pseudo-global-warming (PGW) experiments with high-resolution models have demonstrated their value in projecting thermodynamic changes in sub-daily precipitation extremes under a changing climate scenario (Ban et al., 2015; Prein et al., 2016).

Despite these advances, questions remain about how best to represent sub-grid convective processes such as convection frequency and storm occurrence in coarse-resolution models and climate diagnostics. Convection-permitting model simulations are computationally expensive, limiting the feasibility of producing large multi-year ensembles. Machine learning (ML) offers a promising pathway to address this gap by learning statistical relationships between large-scale environmental variables and sub-grid cloud characteristics. In this study we aim to answer two core scientific questions:

1. *Can ML models predict or represent sub-grid-scale cloud information from coarser-resolution environmental variables?*

2. *What environmental variables most strongly contribute to sub-grid-scale convection according to ML models?*

These questions are central to advancing the application of ML to bridge the scale gap between coarse-resolution climate simulations and unresolved convective processes, with implications for both understanding and modeling weather and climate extremes.

Section 2 describes the datasets, feature engineering, and ML training procedures used in this study. Section 3 presents the performance of the model trained in this study. In Section 4, we evaluate the applicability and generalizability of the model, focusing on how different environmental variables and conditions influence its predictions. Section 5 concludes with a summary of the key findings.

## 2. Data and methodology

The primary objective of this study is to establish relationships between the environmental conditions and sub-grid-scale cloud features. To achieve this, we utilize long-term, convection-permitting regional climate model simulations in which both environmental variables and cloud characteristics are reasonably represented. The four-kilometer long-term regional hydroclimate reanalysis over the contiguous United States (CONUS404; Rasmussen et al. 2023) dataset is selected. It was generated using the Weather Research and Forecasting (WRF) model at 4-km spatial resolution and hourly temporal output frequency. The dataset spans over 40 years (1979–2021) and includes both a current climate control simulation (CTRL) and a pseudo-global-warming (PGW) simulation representing a future climate scenario. The CTRL simulation downscales the ERA5 global reanalysis, providing detailed meteorological variables such as temperature, wind speed, surface pressure, and thermodynamic parameters. Specifically designed to support hydrological modeling and meteorological analysis, the high spatial and temporal resolution of CONUS404 captures mesoscale phenomena and extreme events, offering improved representation of clouds, hydrometeors, and radar reflectivity. This makes it a valuable resource for studying processes and relationships at the weather–climate interface (Rasmussen et al., 2023).

*a. CONUS404 environmental variable selection and convective spatial frequency*

Previous studies utilizing satellite observations have identified key environmental factors useful for distinguishing convective storm modes (Schulte et al., 2024). Building on these findings and variables commonly available in GCMs, we select 16 environmental variables from the CONUS404 output to train the ML models for predicting relationships with the cloud features (Table 1).

| CONUS404 Variable Names | Variable descriptions |
|---|---|
| PWAT | Precipitable water |
| totalVap | Column-integrated water vapor content |
| QVAPOR | Water vapor mixing ratio at the lowest model level |
| PSFC | Surface pressure |
| T2 | Temperature at 2 meters |
| Q2 | Water vapor mixing ratio at 2 meters |
| MLCAPE, MUCAPE | Mixed-layer and most-unstable layer CAPE |
| MLCINH, MUCINH | Mixed-layer and most-unstable layer CINH |
| SBLCL | Surface-based lifting condensation level |
| USHR6, VSHR6 | 0-6 km above-ground-level U, V-component wind shear |
| USHR1, VSHR1 | 0-1 km above-ground-level U, V-component wind shear |
| W | W-component of wind at the lowest model level |

Table 1. List and description of the selected 16 CONUS404 environmental variables used as training features for the ML model.

In addition, we developed an algorithm (Yu, 2022) based on the TRMM-heritage storm mode classification introduced by Houze et al. (2007) and applied it to the simulated radar reflectivity from CONUS404. Convective and stratiform clouds are identified using the masking method of Steiner et al. (1995). For both 40-years of CTRL and PGW simulations, we classified the convective clouds and computed their spatial frequency as the primary sub-grid-scale cloud proxy for ML model label data. This approach enabled us to generate an extensive 40-year database of convective and stratiform cloud classifications, along with storm modes, for both current and future climate simulations.

The native horizontal resolution of the CONUS404 output is 4 km (1015 × 1367 grid points for the entire domain). To approximate the coarse resolution typically used in GCMs, the selected 16 environmental variables are spatially averaged to a 28-km grid (145 × 195 grid points for the entire domain). The spatial frequency of convection is then calculated by computing the fractional coverage of 4-km convective grid points within each 28-km grid cell, referred to here as convective spatial frequency. ML models are then trained to relate the environmental variables to the convective spatial frequency.

### b. Dataset for ML models training

The coarsened environmental variables and the derived convective spatial frequency from the CONUS404 CTRL simulation serve as the input features and target labels, respectively, for the ML training process. Due to the substantial size of the 40-year CONUS404 dataset, we use data from 2018–2021 for training and validation, and data from January to September 2022 for primary testing (Section 3a). To further assess generalizability across different climate regimes, we also conduct additional tests using 12 full years outside of 2018–2022 from both CTRL and PGW simulations (Section 3b).

All hourly data of the 16 selected environmental variables from 2018–2021 are flattened into 16 one-dimensional input features. The corresponding convective spatial frequency is flattened in the same way to serve as the target labels. To improve computational efficiency while preserving the inherent spatial imbalance of convective activity, we reduce the size of the dataset by randomly removing 90% of the data samples. After reduction, each feature array still contains 99,143,460 samples. The data are randomly shuffled and split into 80% for training and 20% for validation. Sensitivity tests (not shown) with varying percentages of data removal show negligible impact on the results, suggesting that the model was trained with a sufficient amount of data to robustly capture the relationship between environmental variables and convective spatial frequency. Finally, the input features are standardized to zero mean and unit variance prior to use in all model training. In contrast, the output, namely the convective spatial frequencies, are not modified. The testing dataset, which was obtained from year 2022 and other years, was not reduced in size.

### c. ML model architecture

We employ a very simple neural network type, namely a multi-layer perceptron (MLP; also known as a dense or fully-connected neural network) to model the relationship between coarsened environmental variables and convective spatial frequency. The MLP is used to approximate the mapping from environmental variables to convection spatial frequencies in a pixel-wise manner (Figure 1). The model architecture is pixel-based, treating each grid point independently based solely on its environmental conditions. The input features do not include any explicit spatial (e.g., latitude or longitude) or temporal information. As a result, the relationships learned by the trained model primarily reflect the intrinsic connections between the 28-km resolution environmental variables and the corresponding convective spatial frequency, rather than being influenced by geographic or time-specific patterns.

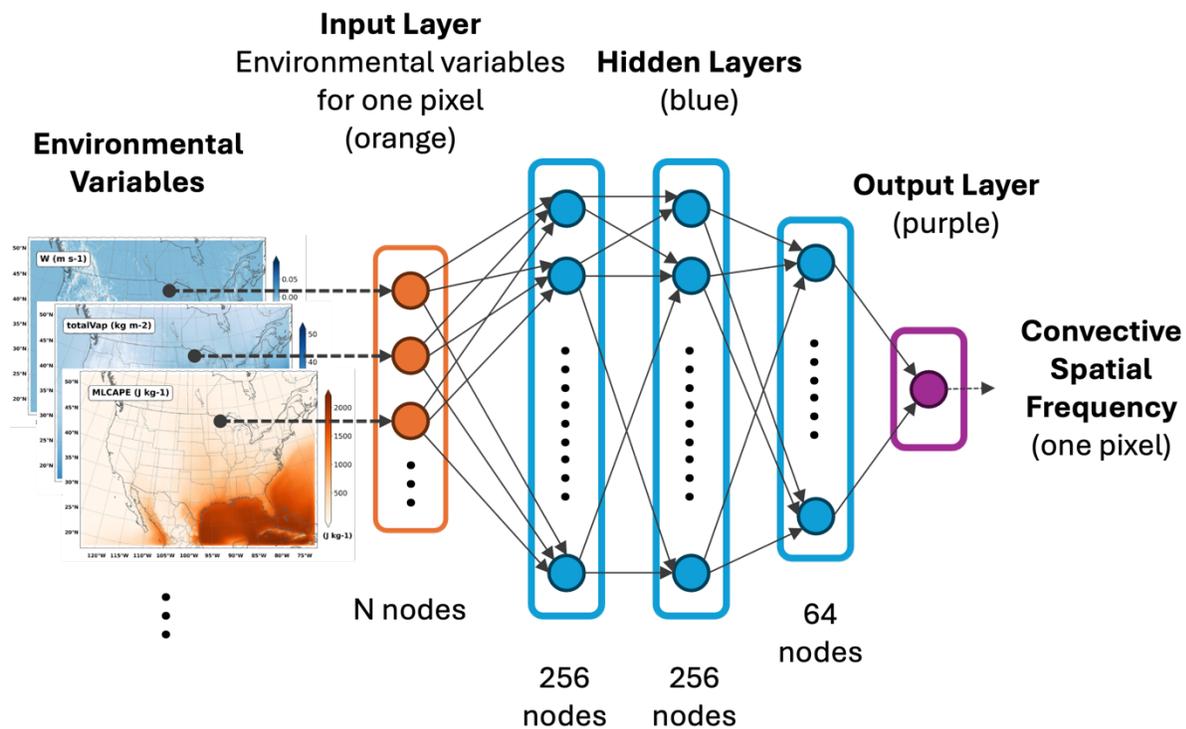

Figure 1. Schematic diagram of the MLP architecture used in this study. Each input node corresponds to an environmental variable at a single grid point. The inputs are passed through three fully connected hidden layers with ReLU activation functions, followed by an output layer predicting convective spatial frequency. The architecture treats each grid point independently, enabling value-to-value predictions without relying on the spatial context.

The motivation for using a simple ML model in this study is twofold:

- *Adaptability and generalization.* Unlike convolutional neural networks, which are highly sensitive to input grid size and spatial structure, the pixel-based model architecture used in this study avoids such limitations by treating each grid point independently. This design enables value-to-value prediction (per pixel) without relying on spatial context, making it more adaptable to resolution changes and suitable for out-of-regime generalization; that is, it is less prone to overfitting.

- *Focus on physical relationships.* The pixel-wise design and exclusion of spatial and temporal information in the input data forces the neural network to focus entirely on modeling direct relationships between the pixel-wise environmental state and the corresponding convective spatial frequency, preventing it from using correlations with temporal or spatial patterns. This fact is particularly advantageous for scientific discovery, specifically for understanding the relationship between environmental conditions and clouds, especially convective characteristics, and for applying the model across diverse climate model grids or under future, previously unseen environmental conditions.

As shown in Figure 1, the MLP consists of four fully connected layers: an input layer followed by three hidden layers with 256, 256, and 64 neurons. Each hidden layer uses a rectified linear unit (ReLU) activation function. The final output layer consists of a single neuron with a linear activation function producing the predicted convective spatial frequency. This architecture is designed to capture complex nonlinear interactions among the features while maintaining computational efficiency.

### d. Linear Regression as Baseline Model

To provide a baseline for comparison, we also train a linear regression (LR) model using the same input features and target label. The LR model is fitted using the standard ordinary least squares method. This baseline model allows us to assess the added value of using a nonlinear multi-layer perceptron by comparing its performance against a simple linear approach, as there is no point in using a neural network model if a linear model is sufficient.

## 3. Model Performance

We use hourly convective frequency data from the CONUS404 CTRL simulation from January to September 2022 as the testing target (ground truth) to evaluate the model performance. We apply the trained models to the hourly environmental variables on the coarsened 28-km grid to generate hourly predictions of convective spatial frequency for the testing period. The predicted values are then compared to the ground truth on a grid-point-by-grid-point basis to evaluate the model performance. As shown in Table 2, the MLP model outperforms the LR baseline in terms of the root-mean-square error (RMSE) and average spatial and temporal correlation coefficients, indicating a more accurate representation of the convective spatial frequency.

We also utilize the structural similarity index measure (SSIM; Wang et al., 2004) to evaluate the model performance. SSIM is a widely used metric in computer vision for quantifying the similarity between two spatial fields or images by jointly assessing their luminance, contrast, and structural information. Unlike point-wise error metrics such as RMSE, which measure only the magnitude of differences at individual grid points, SSIM provides a more holistic evaluation of spatial pattern agreement, making it especially suitable for assessing the structural fidelity of weather variables (Dougherty and Rasmussen, 2021).

In this study, SSIM is computed over patches of 7 × 7 grids (approximately 200 × 200 km with 28-km resolution), allowing us to capture the similarity of convective structures at a meso-to-synoptic scale. The SSIM values range from 0 to 1, where 1 indicates perfect structural similarity between the model prediction and ground truth, and values closer to 0 indicate little to no structural resemblance. The MLP model outperforms the LR model with an overall SSIM value of 0.88 vs. 0.60 as shown in Table 2, which is a notable improvement.

To assess the model robustness, we trained an additional 15 MLP models using different random initializations and data sample reduction configurations. The resulting range of performance metrics demonstrates that the MLP model is robust and consistently provides reasonable predictions of the convective frequency.

|       | RMSE (%)         | Average Spatial Corr. | Average Temporal Corr. | SSIM              |
| ----- | ---------------- | --------------------- | ---------------------- | ----------------- |
| MLP   | 3.377 ± 0.008    | 0.646 ± 0.0016        | 0.580 ± 0.0012         | 0.88 ± 0.0009     |
| LR    | 4.34             | 0.27                  | 0.26                   | 0.60              |

Table 2. RMSE, average spatial and temporal correlation coefficients, and Structural Similarity Image Measure (SSIM), for MLP and LR predictions of convective spatial frequency during January–September 2022, compared with the CONUS404 CTRL simulation ground truth. The range of MLP values represents the results of 16 trained MLP models.

*a. MLP and LR model performance*

To further evaluate model performance beyond general metrics, it is crucial to examine the spatial distribution of convective clouds, which is a key consideration in weather and climate research. Understanding where the most intense convective activity occurs provides insight into the model's ability to represent physically meaningful patterns. Figure 2 shows the ground truth convective spatial frequency over a 28-km resolution grid, averaged from January to September 2022, based on the CONUS404 CTRL simulation. The CONUS404 model effectively captures prominent convective hotspots, including oceanic regions such as the storm track and Gulf Coast, as well as land regions like the Great Plains, Rocky Mountains, and Sierra Madre mountain range in Mexico.

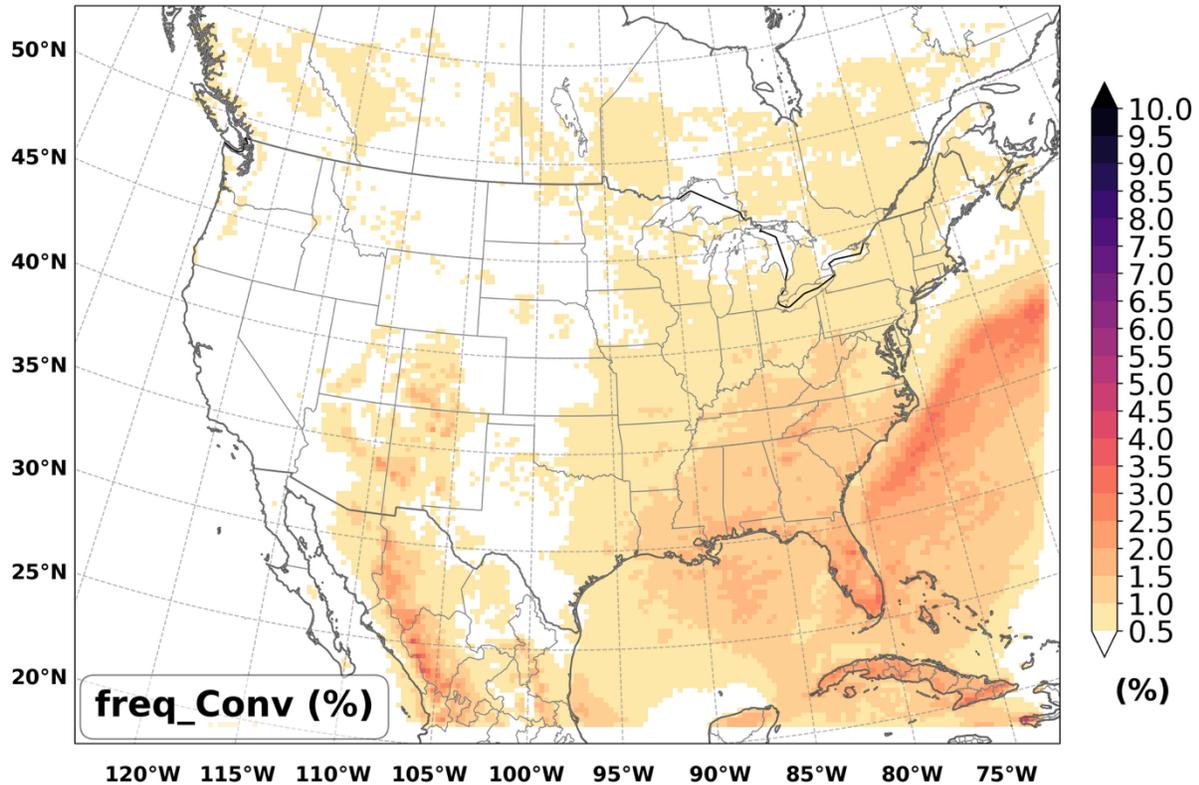

Figure 2. Average spatial frequency of convective clouds from the CONUS404 CTRL simulation during January–September 2022. The grid resolution is 28-km.

The MLP model prediction (Figure 3a) successfully captures the prominent convective hotspots identified in the ground truth, relying solely on pixel-based environmental information. The difference map (Figure 3b) indicates that the MLP predictions exhibit only slight biases relative to the ground truth. In contrast, the LR model trained with the same input features is only able to represent the broad continental-scale pattern of convection (Figure 3c). Its difference map reveals substantially larger biases than the MLP, with pronounced overestimation along the Southern Plains and underestimation over the Rocky Mountains, Sierra Madre mountain range, and coastal regions, including the storm track (Figure 3d).

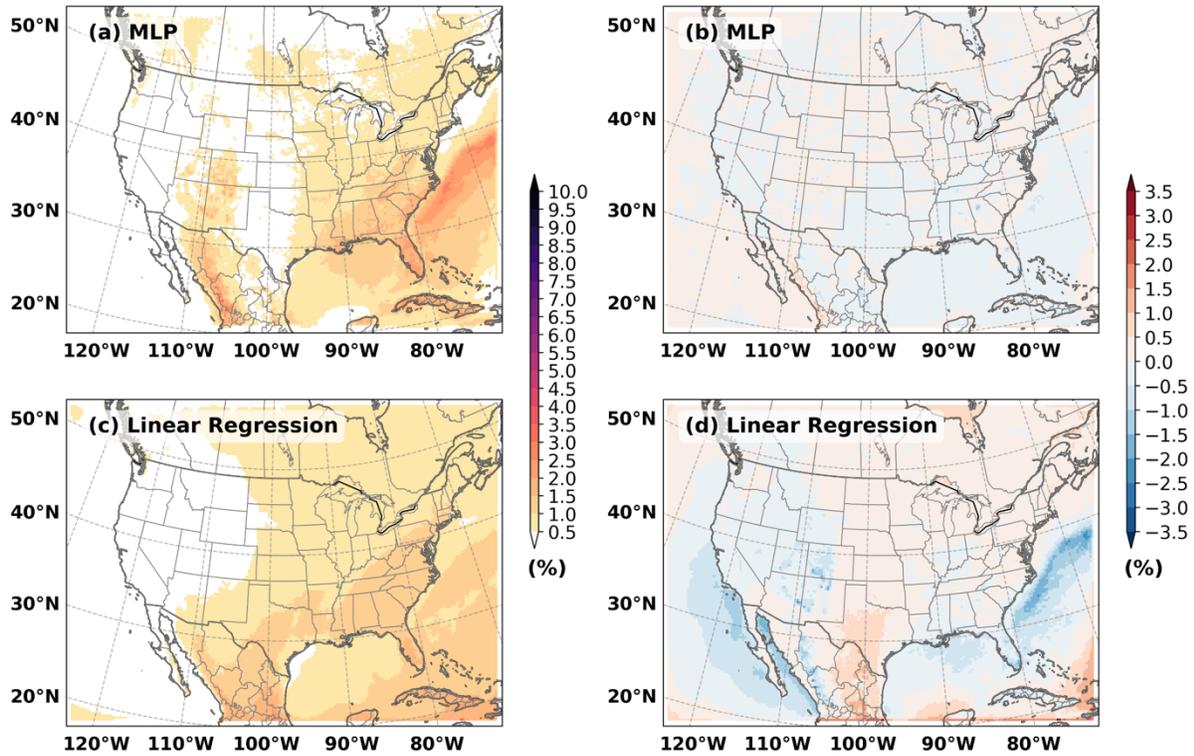

Figure 3. Left: Average MLP (a) and LR (c) model predictions of convective frequency during January–September 2022. Right: Differences between model predictions (b: MLP; d: LR) and ground truth.

The MLP model performance is also evaluated for extreme weather events to assess whether the model can capture the timing and spatial patterns of convective activity. Figure 4 shows the result during a multi-day severe weather outbreak across the southern United States in late March 2022. The event was driven by an eastward-moving midlatitude cyclone that triggered multiple rounds of severe and tornadic storms. Figure 4 shows the convective spatial frequency from the CONUS404 CTRL simulation and MLP prediction for the evening of March 21 (00Z on March 22, 2022), when numerous tornadoes impacted East Texas, including areas near the Dallas and Austin metro regions. The CONUS404 CTRL simulation captures the event well as the coarsened 28-km resolution convective spatial frequency clearly depicts the bow-echo MCS structure over Texas (Figure 4a). A zoom-in reveals how the coarsened 28-km grid represents the convective structure as seen in the native 4-km resolution (Figure 4c).

The MLP model also reproduces this event reasonably well compared to the ground truth. Although the MLP prediction exhibits a broader spatial extent due to the smoothing tendency of neural networks, the bow-shaped convective pattern is preserved (Figure 4e), and regions of higher predicted convective frequency closely align with the native 4-km distribution (Figure 4f), highlighting the MLP model's ability to capture extreme values.

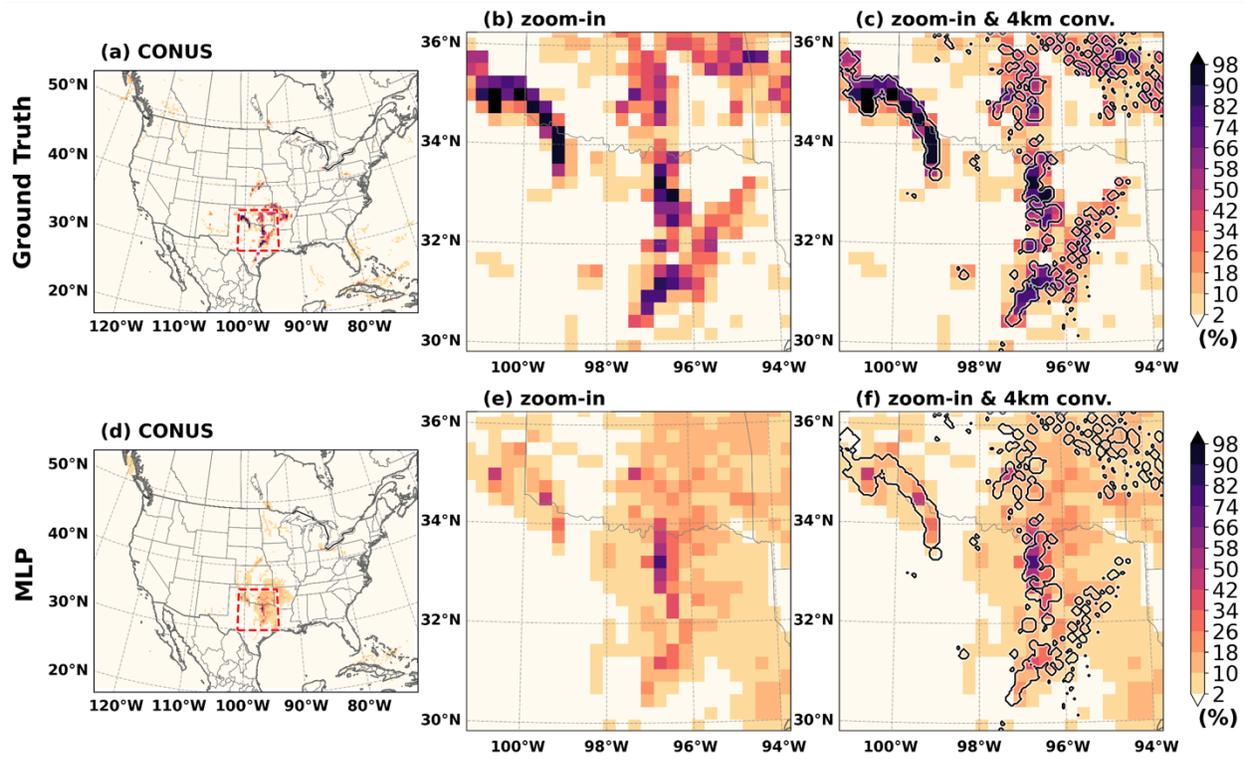

Figure 4. Convective spatial frequency at 28-km grid resolution derived from the CONUS404 CTRL simulation (a, b, c) and predicted by the MLP model (d, e, f) at 00Z on March 22, 2022. The red dashed boxes in the left panels indicate the region shown in greater detail in the middle and right panels. The colored grids in the middle and right columns are identical (b and c; e and f), except for the highlighted black contours in the right panels which denote convective cloud areas at the native 4-km resolution from the CONUS404 CTRL simulation (ground truth).

Figure 5 presents an additional evaluation of convective activity across the CONUS domain. At this hour, the domain-averaged SSIM reaches its lowest value during the testing period (~0.72), due to widespread convection. Although this is the minimum domain-averaged SSIM

File generated with AMS Word template 2.0

throughout the primary testing data, values above 0.7 are generally regarded as indicating good structural similarity in image analysis applications, underscoring that the MLP model maintains skill even under the challenging convective conditions. Focusing on mountainous regions (Figure 5b, c, e, f), the MLP model tends to smooth areas of higher convective spatial frequency and occasionally misses extreme convective peaks and lacks clearly defined structures. Despite demonstrating strong performance at the domain scale and capturing the general locations of active convection, the model still requires improvement in the representation of fine-scale convective details. Notably, the MLP was trained solely on flattened, randomly downsampled 28-km environmental variables without any spatial, temporal, or geographic context, highlighting both the model's strengths and its current limitations.

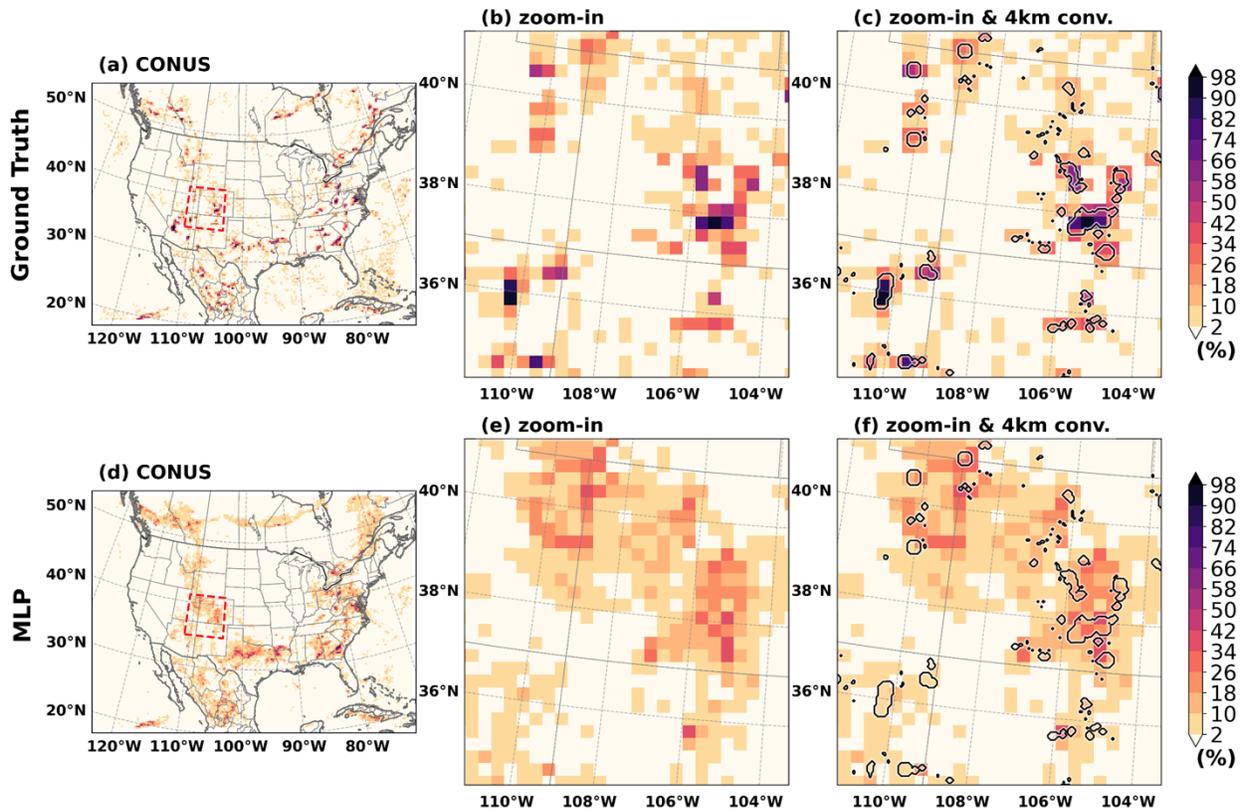

Figure 5. Like Figure 4 except for 02Z on August 22, 2022

File generated with AMS Word template 2.0

The MLP model also captures the temporal variations of convective activity. Figure 6a and b shows the daily average domain-mean convective frequency time series from January to September 2022 for the ground truth, MLP, and LR predictions, and the corresponding difference time series. The MLP predictions align closely with the ground truth, yielding a temporal correlation coefficient of approximately 0.95. Although the LR model generally reproduces the seasonal cycle with a higher convective frequency during the warmer months, it exhibits larger biases throughout the year, characterized by underestimation until May and overestimation from June onward, resulting in a lower correlation coefficient of 0.75.

Figure 6c shows the daily-averaged domain-mean SSIM time series for the MLP and LR models during the testing period. As shown, the MLP model outperforms the LR model in reproducing the structural characteristics of the convective frequency, with SSIM values consistently closer to 1 and exhibiting more stable variation. Both models show a decline in the SSIM values during the warmer months, with minimum values occurring in August, suggesting reduced ability to represent accurate convective spatial distributions when convective activity is more prominent. Notably, the MLP SSIM decreases modestly from approximately 0.9 to 0.8, whereas the LR SSIM decreases more sharply from approximately 0.7 to below 0.5.

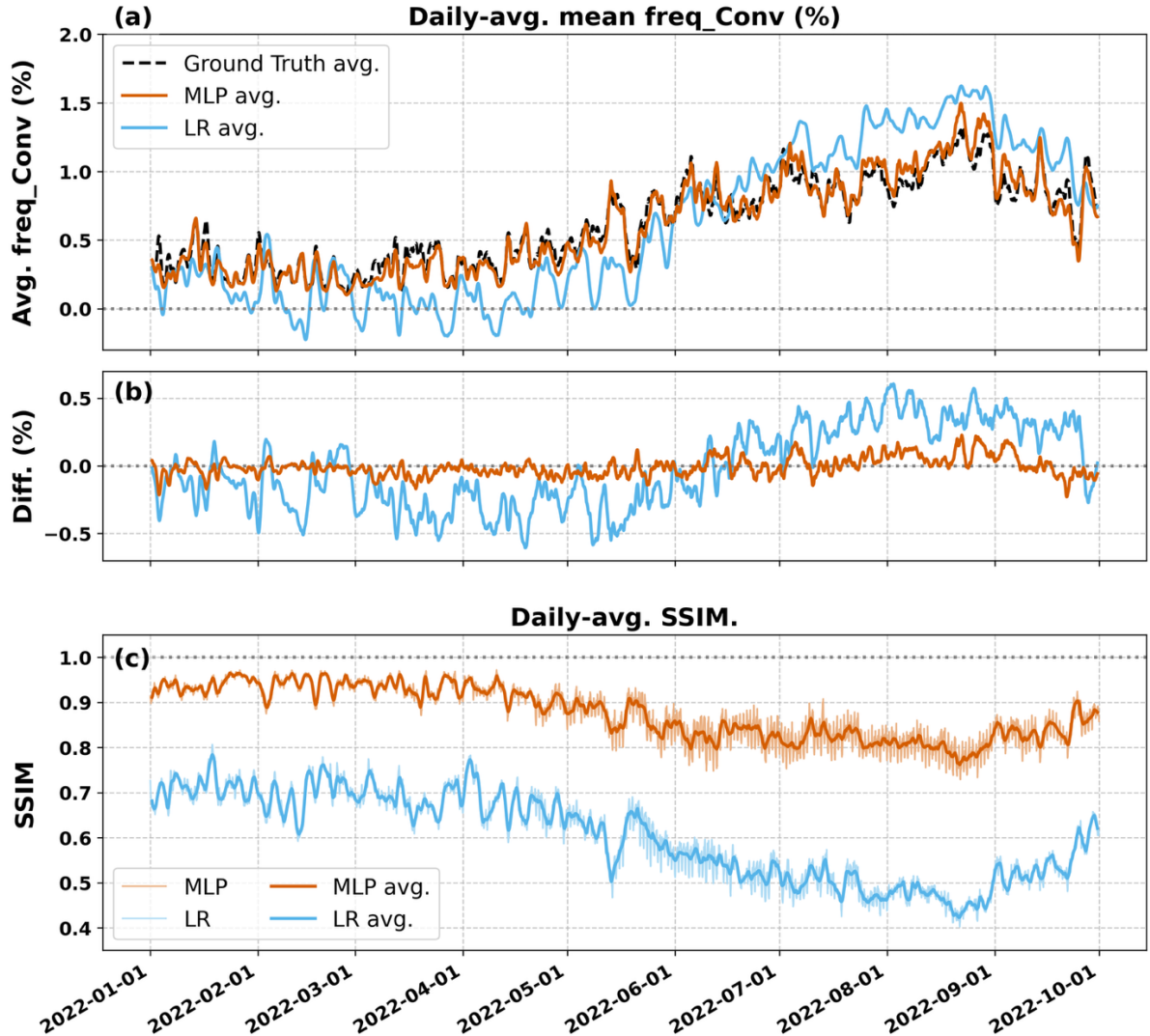

Figure 6. (a): Daily average domain-mean convective spatial frequency from January to September 2022 for the ground truth (dotted black), MLP (orange), and LR (blue) predictions. (b): the differences between the MLP and LR predictions and the ground truth. (c): the daily average SSIM of the MLP and LR predictions relative to the ground truth (black dotted line equals to 1), with hourly SSIM indicated by lighter colors.

The MLP model prediction also captures the diurnal cycle of convection frequency reasonably well, outperforming the LR model (Figure 7a). The afternoon peak in convective spatial frequency is reasonably reproduced, with a peak delay of approximately two hours and an average underestimation bias of approximately 0.1%. A slight increase in nocturnal to early morning convection is also represented, although with a modest overestimation. Notably, the MLP model is applied to hourly environmental variables, but the input data contain no explicit temporal information and are flattened and treated in a purely pixel-based manner. This suggests that the MLP model's ability to relate environmental variables to convective frequency enables it to reproduce key features of the observed diurnal cycle. This includes the afternoon peak in convective activity across the eastern U.S., which is commonly associated with solar heating and daytime destabilization (Parker & Ahijevych, 2007), as well as the nocturnal peak over the central Great Plains, linked to mesoscale convective systems (MCSs) maintained by processes such as low-level jets and elevated CAPE (Geerts et al., 2017; Blake et al. 2017). Such features are challenging to capture without explicit temporal or spatial context, underscoring the ability of the MLP model to infer these relationships from environmental inputs alone.

The diurnal cycle of the standard deviation (Figure 7b) shows that the MLP model tends to smooth the predictions, leading to a reduced variance compared to the ground truth. In contrast, the LR model lacks the capacity to capture interactions between environmental drivers and convection and struggles to represent both the magnitude and variability of the diurnal cycle standard deviation.

Importantly, representing the diurnal cycle of convection remains a persistent challenge for GCMs, particularly because of the coarse resolution and parameterized physics (Song et al. 2024; Zhang et al. 2024). This analysis highlights the potential of machine learning techniques to retrieve realistic convective diurnal cycles from coarse-resolution environmental variables, offering a promising approach to bridge the scale gap between large-scale climate models and sub-grid convective processes.

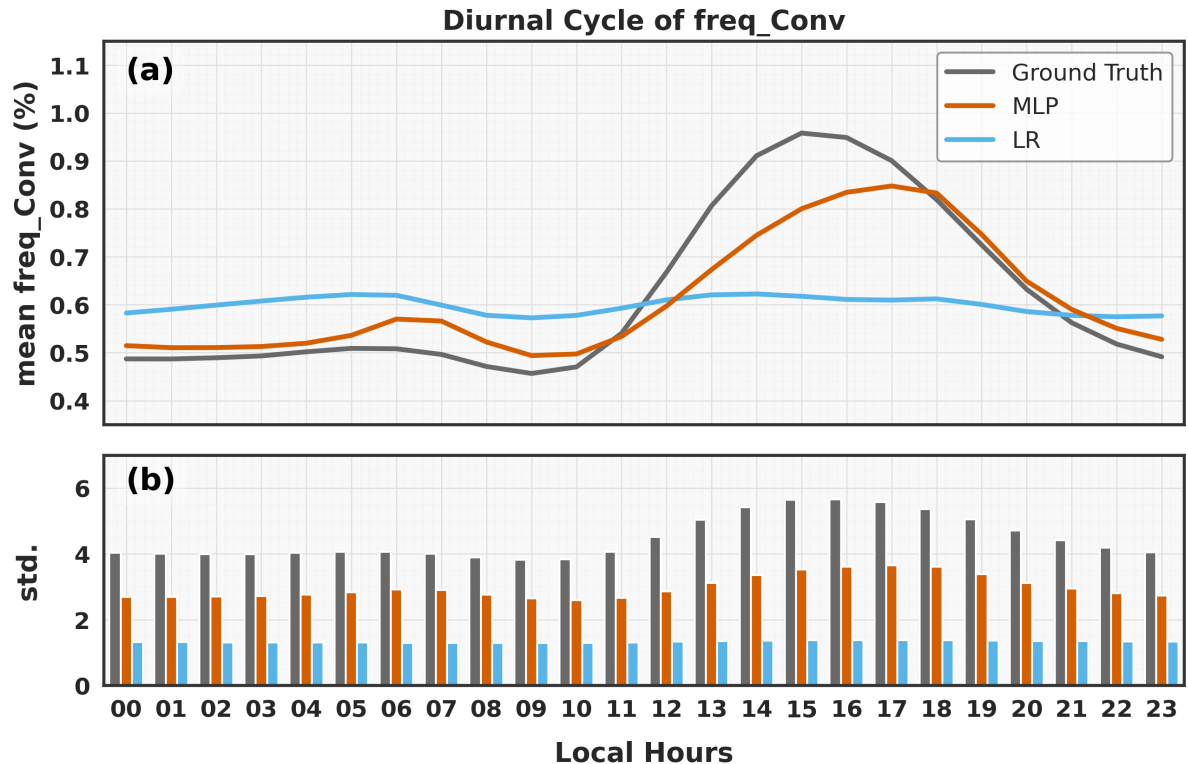

Figure 7. (a): Average diurnal cycle of convective spatial frequency from January to September 2022 for the ground truth, MLP predictions, and LR predictions. (b): the diurnal cycles of the standard deviations.

b. *MLP model applicability in different climate regimes and grid resolutions*

The CONUS404 simulation dataset includes two 40-year sets of simulations: a CTRL run representing the current climate and a PGW run representing a warmer, moister future climate scenario. From the CTRL simulation, we select 12 years outside the MLP model's training period (2018–2021) that capture a range of global-scale climate variability to evaluate how well the MLP model predicts convective spatial frequency under different climate conditions (Table 3). The corresponding 12 years from the PGW simulations are also used to assess the model performance in a warmer, moisture-rich environment.

| Year | Description |
|------|-------------|
| 1982 | Strong El Niño; warm winter; beginning of a major ENSO event. |
| 1988 | Strong La Niña; widespread U.S. drought and heat wave. |
| 1993 | Neutral ENSO; Midwest flooding (Great Flood of 1993). |
| 1997 | Very strong El Niño onset; wet West Coast winter. |
| 1998 | Strongest El Niño peak; warmest global year at the time; winter storms. |
| 1999 | Strong La Niña; cooler, wetter northern U.S. |
| 2005 | Neutral ENSO; record-breaking Atlantic hurricane season (Katrina, Rita, Wilma). |
| 2011 | Strong La Niña; severe tornado outbreaks, Texas drought. |
| 2012 | Neutral ENSO; extreme drought and record heat across the central U.S. |
| 2013 | ENSO-neutral; generally average year. |
| 2014 | El Niño onset; intensifying California drought. |
| 2015 | Strong El Niño; warm winter; wet southern U.S., dry northern U.S. |

Table 3. List of the selected years and general climate descriptions over the 40-years CONUS404 simulations.

Figure 8 shows the annual variations in the daily average SSIM values for each selected year in the CTRL simulation as scattered points, colored by the corresponding daily average domain-mean convective spatial frequency. The solid lines depict the mean annual variations across all selected years for both the CTRL (black) and PGW (blue) simulations. The mean annual variation in SSIM in the CTRL years is comparable to the testing period (January–September 2022; Figure 6), with overall SSIM values greater than 0.8 and individual years showing consistent variability within ±0.05 throughout the year. The mean SSIM for the MLP predictions on the PGW simulations is nearly identical to that of the CTRL run, with SSIM value differences on the order of ~0.1.

Annual variations further reveal a decline in MLP performance during warmer months with a higher convective spatial frequency. The colored dots in Figure 8 illustrate that days with higher domain-mean convective frequency tend to correspond to lower SSIM values, clustering toward the lower portion of the distribution. Although the MLP performance is slightly reduced during periods of more active convection, the minimum SSIM average remains greater than 0.75, and as shown in Figure 4 and Figure 5, the model effectively captures regions with high convective frequency compared to the native 4-km resolution convective grids, indicating reliable overall performance.

Note that this study does not aim to evaluate the MLP performance for specific years or explore links to specific climate indices such as El Niño-Southern Oscillation (ENSO). Although these are important open questions, they are beyond the scope of this study. Instead, the results demonstrate that a pixel-based MLP model trained on environmental variables from a limited period can robustly represent sub-grid-scale convection proxies across a range of climate regimes, supporting its applicability in coarse-resolution climate model simulations and future climate projections.

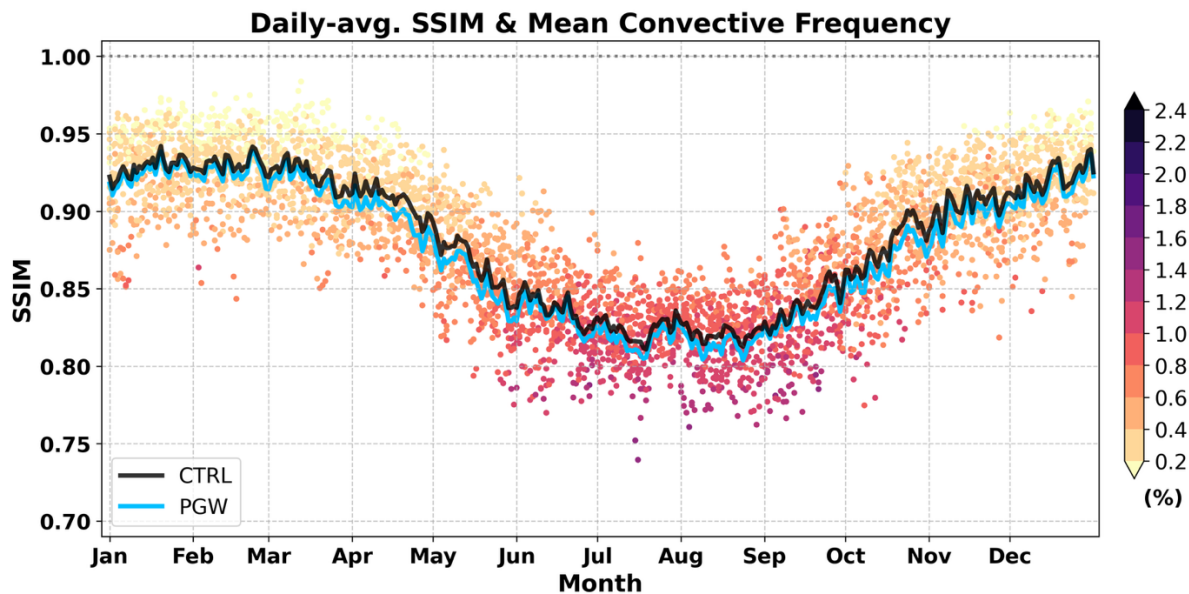

Figure 8. Mean daily average SSIM across all selected years of climate variability for MLP predictions of the CONUS404 CTRL (black) and PGW (blue) simulations. Colored dots indicate

the distribution of SSIM values for individual years in the CTRL simulation, with the shading of each dot representing the domain mean convective frequency for that day.

As part of our overarching goal to explore the applicability of sub-grid-scale cloud proxies in GCMs, we investigate how the MLP model trained using 28-km resolution data performs when applied to even coarser 100-km resolution environmental variables. Figure 9a shows the mean convective spatial frequency at 100-km resolution during the testing period from CONUS404. When the trained MLP model is applied to 100-km resolution environment inputs, it successfully reproduces the spatial patterns of convective frequency, capturing key convection hotspots over land and topography, including the Great Plains, the Rockies, and the Sierra Madre; and over the coastal and oceanic regions, such as the storm track along the East Coast (Figure 9b). The differences from the ground truth are generally small, with overall biases less than 0.5% and only slight underestimation over the storm track (Figure 9c).

These results demonstrate that even when applied at a different spatial resolution, the MLP model effectively learns and preserves the interactions among environmental variables to represent the convective spatial frequency. This capability is a key advantage of the pixel-based MLP architecture which, unlike convolution-based models, is inherently resolution-independent. This finding provides confidence in the applicability of the MLP model to traditional coarse-resolution GCMs.

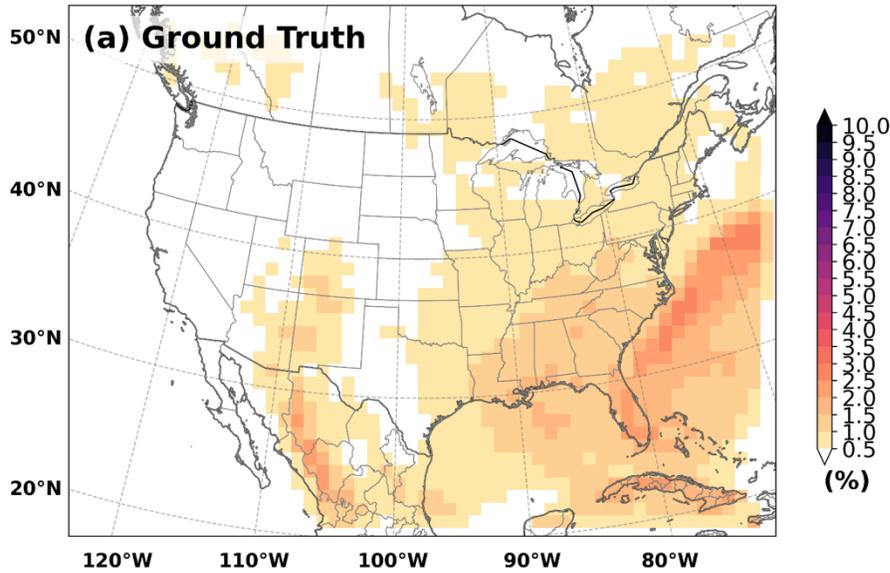

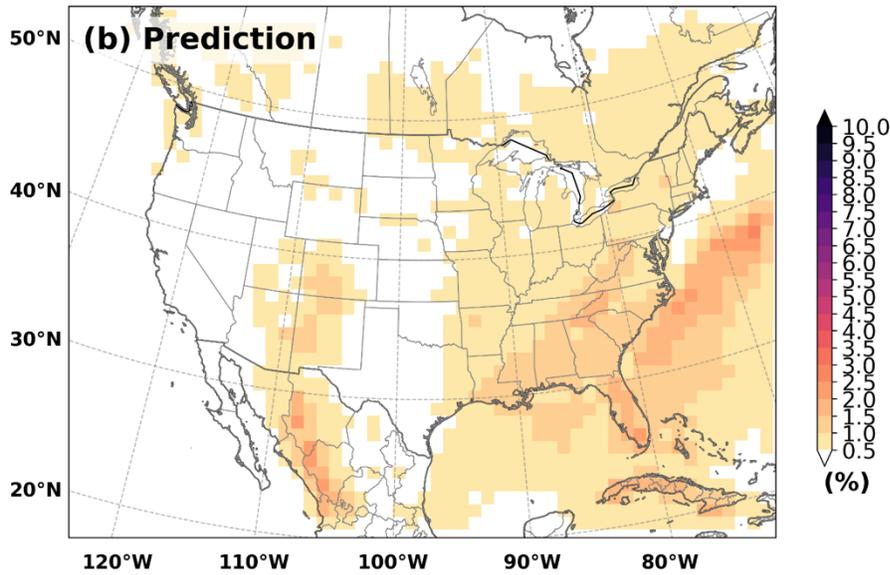

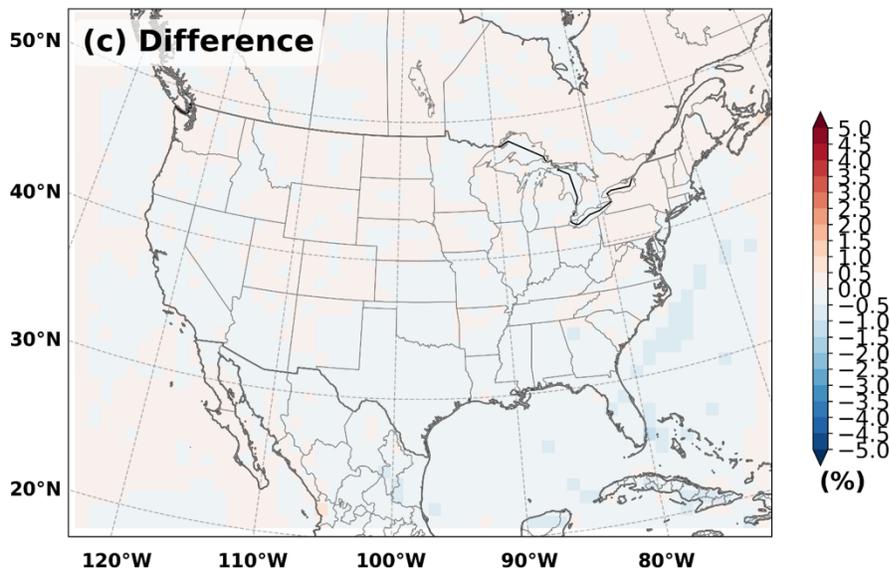

23File generated with AMS Word template 2.0

Figure 9. Average convective spatial frequency during January–September 2022 at 100-km grid resolution for the CONUS404 CTRL simulation (ground truth; a) and MLP model prediction (b). Panel (c) shows the difference between the MLP prediction and ground truth.

## 4. Interpretation for MLP Model

While the trained MLP model is capable of reasonably predicting the sub-grid-scale convective cloud spatial frequency, we also seek to address the second key question: *What environmental variables most strongly contribute to sub-grid-scale convection according to ML models?* To investigate this, we conducted two sets of experiments using different model configurations.

### a. Regional generalizability for MLP model

In the first set of model configuration experiments, we aim to evaluate the generalizability of the MLP model across different geographic regions within the CONUS domain. Convective processes vary regionally across the CONUS, leading to distinct interactions between environmental conditions and convection. By training the model using data restricted to specific regions, we assess whether the model can capture these regional differences and accurately predict convective behavior beyond the training domain. This approach also helps identify regions where the model performance is most sensitive to environmental characteristics and whether the MLP architecture can represent different convective processes across diverse geographic settings. In addition, this type of regional exclusion analysis is feasible with pixel-based models but not with convolutional neural network (CNN) architectures, which rely on spatial structures and neighborhood information to make predictions. This provides insights into the spatial robustness of the pixel-based MLP architecture that treats each grid point independently.

First, we divide the CONUS domain into four regions: north and south of 30°N (labeled as North and South), and east and west of 100°W (labeled as East and West). This separation is informed by the distribution of prominent convective hotspots across the CONUS, as shown in Figure 2. Each region encompasses distinct convective regimes and associated geographical environments. For each region, an independent MLP model was trained using all 16 input features but limited to data within the respective region. The performance in SSIM values of these regionally trained models during the warm months of the testing period (May–September 2022) is compared to that of the model trained on the full CONUS domain, as shown in Figure 10.

The results show that each regionally trained model performs comparably to the full-domain model when applied within its own region. However, the performance slightly degrades when applied to the opposite regions. For example, the Northern-trained model performs similarly to the full-domain model when predicting over the Northern region (Figure 10a, solid line), whereas the Southern-trained model exhibits poorer performance over the North, with SSIM values lower by approximately 0.03 (Figure 10a, dashed line). The largest performance decline occurs when applying the Eastern-trained model to the Western region (Figure 10d; ~0.05 SSIM drop in August). These findings indicate that although the MLP model effectively captures the interactions among environmental variables within a region, its performance is sensitive to the distinct environmental conditions of different regions. This suggests that the MLP model benefits from exposure to region-specific environmental information during training to better predict sub-grid convective frequency, as different regional characteristics influence the relationships between environmental variables and convection.

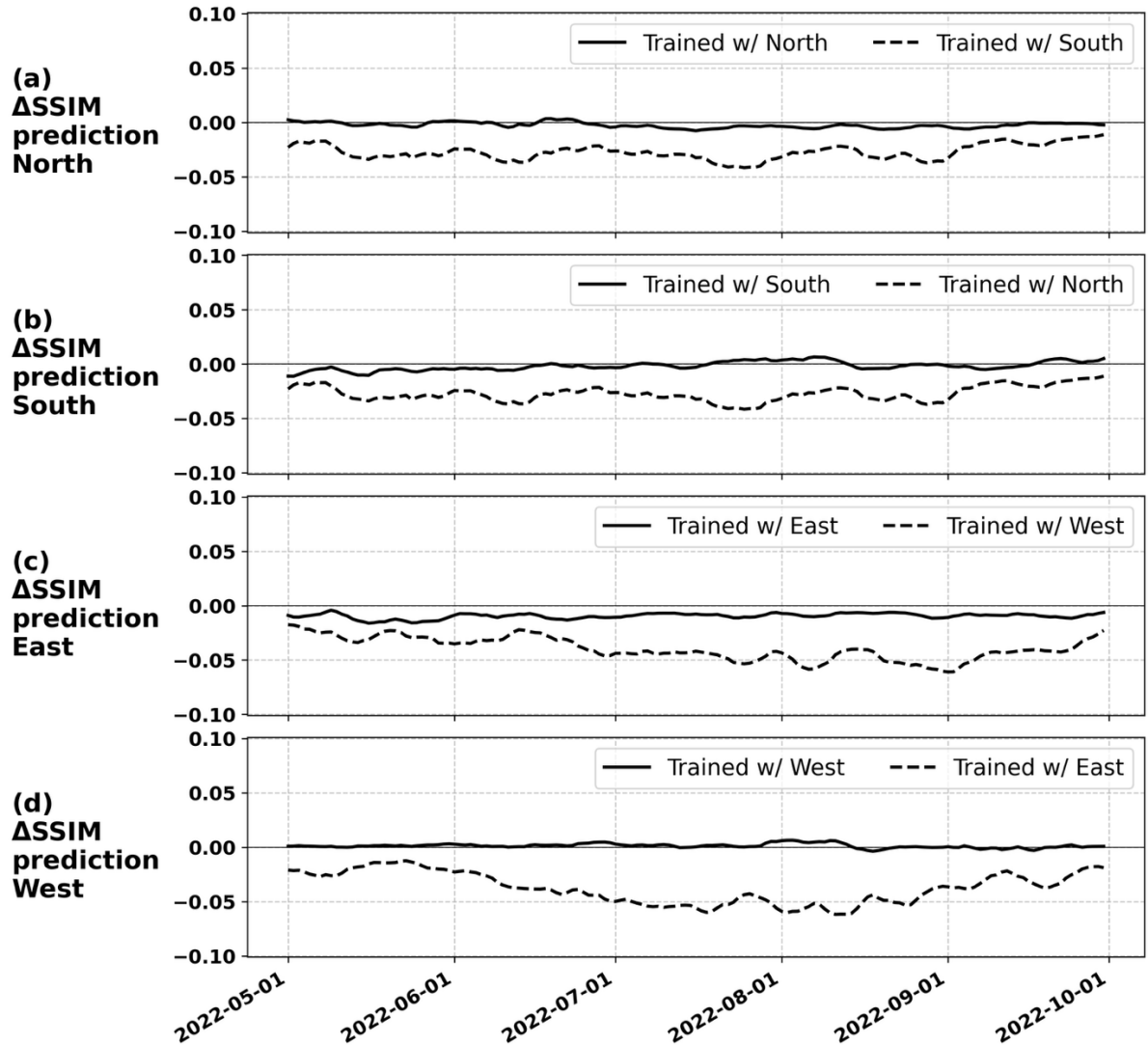

Figure 10. Differences in SSIM values for regional MLP model predictions across different geographic domains: north and south of 30°N (a, b), and east and west of 100°W (c, d), evaluated during the warm season of the testing period (May–September 2022). In each panel, the solid line represents the SSIM values for predictions made within the region where the model was trained, whereas the dashed line shows the SSIM values for predictions made outside the training region (i.e., using the model trained on the opposite region).

Figure 11 and Figure 12 illustrate the regions within the CONUS where the MLP model's prediction is more sensitive and demonstrates reduced prediction accuracy. For the models trained using data from the northern and southern regions separately, Figure 11a and Figure 11d show that the regionally trained models perform reasonably well within their respective training domains, with only minor biases. The model trained on northern data slightly overestimates the convective frequency along the eastern coastal storm track (Figure 11a), likely because of the limited information on warm ocean interactions with convection.

In contrast, the model trained on the southern region, which effectively represents the Gulf Coast and subtropical environments in Figure 11d, struggles when applied to the northern region (Figure 9b). It substantially overestimates the convective frequency over the Rocky Mountains, likely because it failed to capture mid-latitude mountain convective processes such as the mountain–plains solenoid, which is characteristic of broad heating over large mountain ranges (Bao and Zhang 2013).

Similarly, Figure 11c shows that the model trained solely on northern data exhibits large biases over the Sierra Madre mountains in the south, shifting the convective frequency westward toward the Pacific coast of Mexico. This misrepresentation suggests the model's inability to capture the key features of the North American Monsoon, which dominates the Sierra Madre region during the warm season (Gochis et al. 2005; Boos and Pascale 2021).

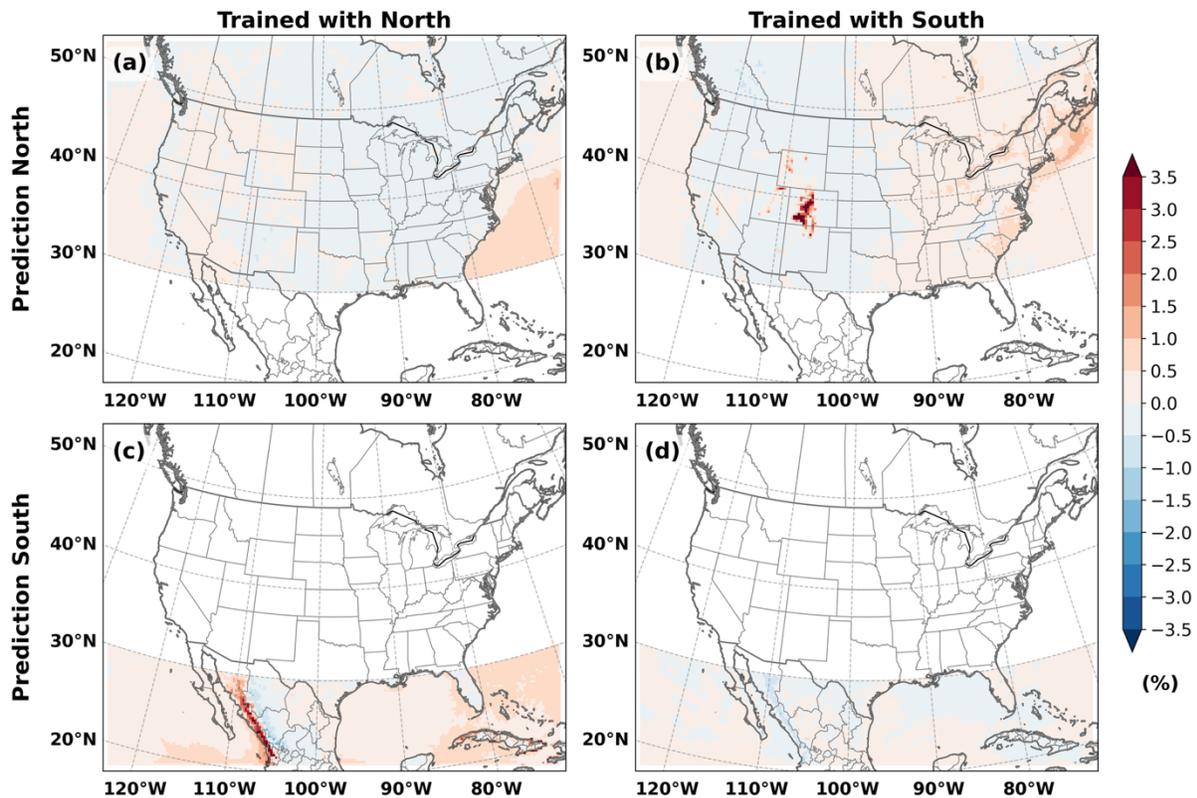

File generated with AMS Word template 2.0

Figure 11. Regional differences in the mean convective spatial frequency during the warm season of the testing period (May–September 2022) for MLP models trained and evaluated over different latitude bands. (a) MLP model trained with data north of 30°N and applied to the northern region; (b) model trained with southern data and applied to the northern region; (c) model trained with northern data and applied to the southern region; and (d) model trained with southern data and applied to the southern region.

Figure 12 presents the results for the models trained using data from the eastern and western regions of the CONUS. Similar to the northern and southern experiments, the regionally trained models exhibit smaller biases when applied within the same training domains (Figure 12a and Figure 12d). However, when applied outside the training regions, clear deficiencies emerge. The model trained over the western CONUS, which lacks exposure to warm coastal waters and associated convective environments, underestimates convective activity along the southeastern U.S. coast, the storm track, and the Caribbean region (Figure 12b). This highlights the importance of warm oceanic environments in shaping convective frequency over the eastern coastal region, which is absent in the western training data.

Conversely, the model trained on eastern CONUS data, which lacks representation of significant mountainous regions, exhibits pronounced positive biases over the Sierra Madre and negative biases over the Rocky Mountains (Figure 12c). This suggests that without mountainous environments in the training data, the model cannot capture the distinct convective processes associated with orographic influences. The biases over the Sierra Madre and Rockies are of opposite signs, further implying the differing convection patterns and the roles of mountainous environments between the subtropics and mid-latitudes (Kodama and Tamaoki 2002).

These results reinforce the notion that mountainous and oceanic environments introduce unique environmental–convection interactions. Therefore, an accurate representation of the sub-grid-scale convective frequency requires geographically diverse training data that capture the full range of environmental regimes, including coastal, oceanic, and mountainous influences.

File generated with AMS Word template 2.0

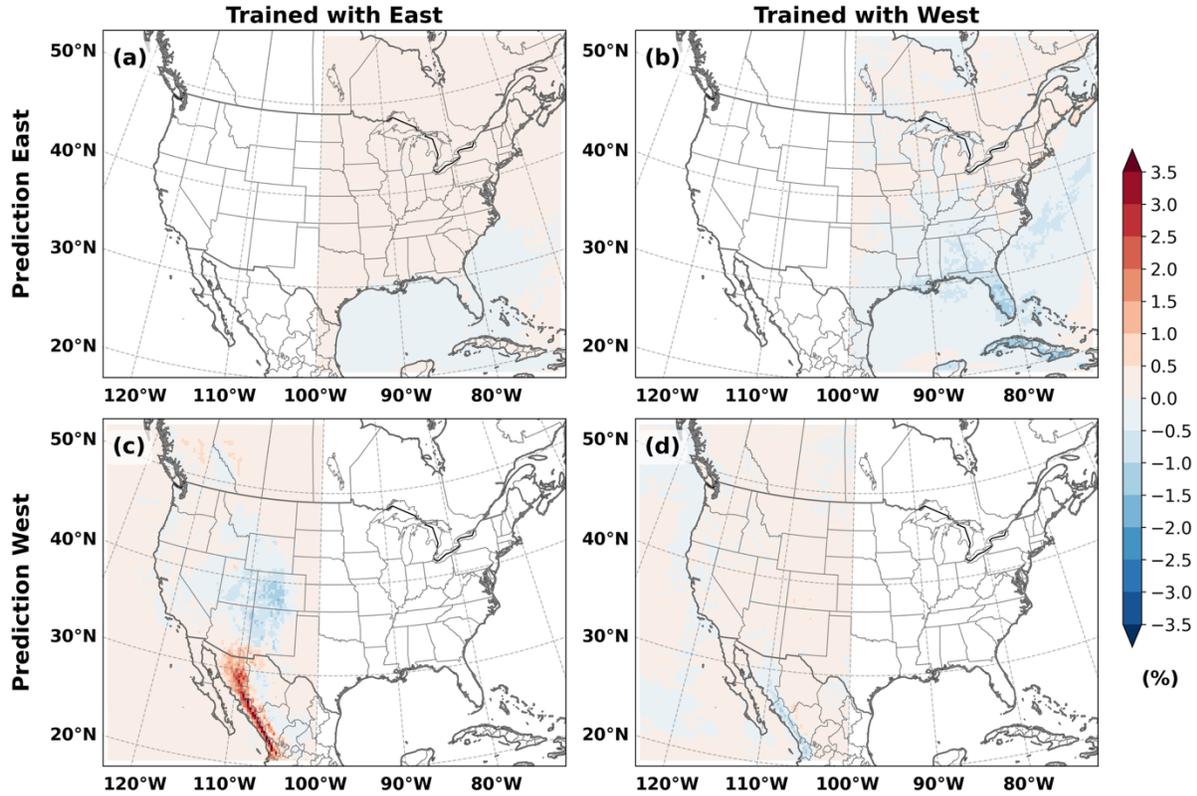

Figure 12. Similar to Figure 11 except for (a) MLP model trained with data east of 100°W and applied to the eastern region; (b) model trained with western data and applied to the eastern region; (c) model trained with eastern data and applied to the western region; and (d) model trained with western data and applied to the western region.

## b. *Feature ablation study for MLP model*

To investigate how the MLP model learns from environmental inputs and which variables most strongly influence performance, we conducted an ablation study by systematically removing groups of input features during model training. The model configurations and corresponding results are shown in Figure 13. Focusing on the warm season of the testing period (May–September 2022), we evaluated how the SSIM values change relative to the full model trained with all 16 features (labeled as feat_16 in Figure 13). This comparison provides insight into which groups of environmental variables are most critical for predicting convective spatial frequency, and to what extent input dimensionality can be reduced while maintaining reasonable accuracy in representing sub-grid-scale convective proxies.

As shown in Figure 13, all ablated models exhibit reduced performance compared to the full model, with the most substantial declines occurring in July and August when convective activity is more prominent. The impact of feature removal varies across configurations, indicating different levels of influence among environmental variables. Notably, although wind shear and column-integrated moisture, including precipitable water (PWAT) and total column water vapor (totalVap), are widely recognized controls on convection (Doswell et al., 1996; Peters et al. 2022), removing only wind shear (model 01 in Figure 13) or column moisture (model 02) results in a relatively small decrease (~0.02) in the SSIM of convective spatial frequency, without significant seasonal differences. This suggests that the model may compensate for their absence by using other correlated features. However, when all moisture-related variables are removed (model 03), the performance declines more substantially than model 02, with the SSIM decreasing by up to ~0.08. Conversely, when only moisture variables are retained (model 08, the paired opposite of model 03), performance degrades sharply with a peak SSIM reduction of ~0.25 in August, approaching the performance of the baseline LR model. This sharp decline likely reflects a lack of sufficient feature interactions, which limits the model's ability to capture the complex interacting processes governing convection.

The next ablation pair examines the impact of removing versus retaining the surface and lowest model level variables. Model 04 is trained without surface and near-surface information, whereas model 07 retains only these variables. Removing the surface and lowest model level variables (model 04) results in slightly worse performance than removing moisture variables (model 03). In contrast, training with only surface and lowest model level variables (model 07) yields better performance than training with only moisture variables (model 08), particularly from mid-June onward when convective activity becomes more widespread. This suggests that surface interactions contribute more significantly to the prediction of convective frequency than moisture fields alone.

Finally, models 05 and 06 focus on the role of thermodynamic indices, including the surface temperature, lifting condensation level (LCL), convective available potential energy (CAPE), and convective inhibition (CINH). Model 05 excludes these indices, whereas model 06 retains only them. These two models show the most similar performance among the ablation pairs, revealing two points: (1) thermodynamic indices are among the most critical environmental variable groups. Among models 02–05, removing thermodynamic indices degrades the model performance the most; and (2) owing to their importance, even a model trained solely on thermodynamic indices (model 06) performs better than models trained solely on either moisture fields or surface variables (models 07 and 08) in predicting convective spatial frequency.

One additional point concerns wind shear. Although the initial ablation results (model 01) suggest that wind shear alone has a limited impact on convective spatial frequency, ablating wind shear from models 02–05 leads to further performance declines (not shown), particularly for models 04 and 05, whose performances then approach those of models 06 and 07, respectively. This indicates that although wind shear may not strongly impact convective spatial frequency prediction by itself, its interactions with other variables are important for capturing convection-related patterns.

Conducting a comprehensive ablation study for all possible feature combinations is impractical. However, by applying domain knowledge to systematically group and remove relevant environmental variables, we gain insight into how the MLP model learns and leverages input information. These results highlight that thermodynamic indices are more critical for representing convective spatial frequency than moisture or surface variables alone. Wind shear, while less impactful on its own, influences the prediction through interactions with other environmental variables.

In summary, the ablation study emphasizes the importance of interactions among environmental variables for capturing sub-grid-scale convection, reaffirming the advantage of using MLP architectures for this task. Furthermore, the interactions among different feature groups reinforce the need to explore these relationships in great depth. In the future, we plan to apply interpretable machine learning techniques to better understand these complex dependencies and improve model transparency.

File generated with AMS Word template 2.0

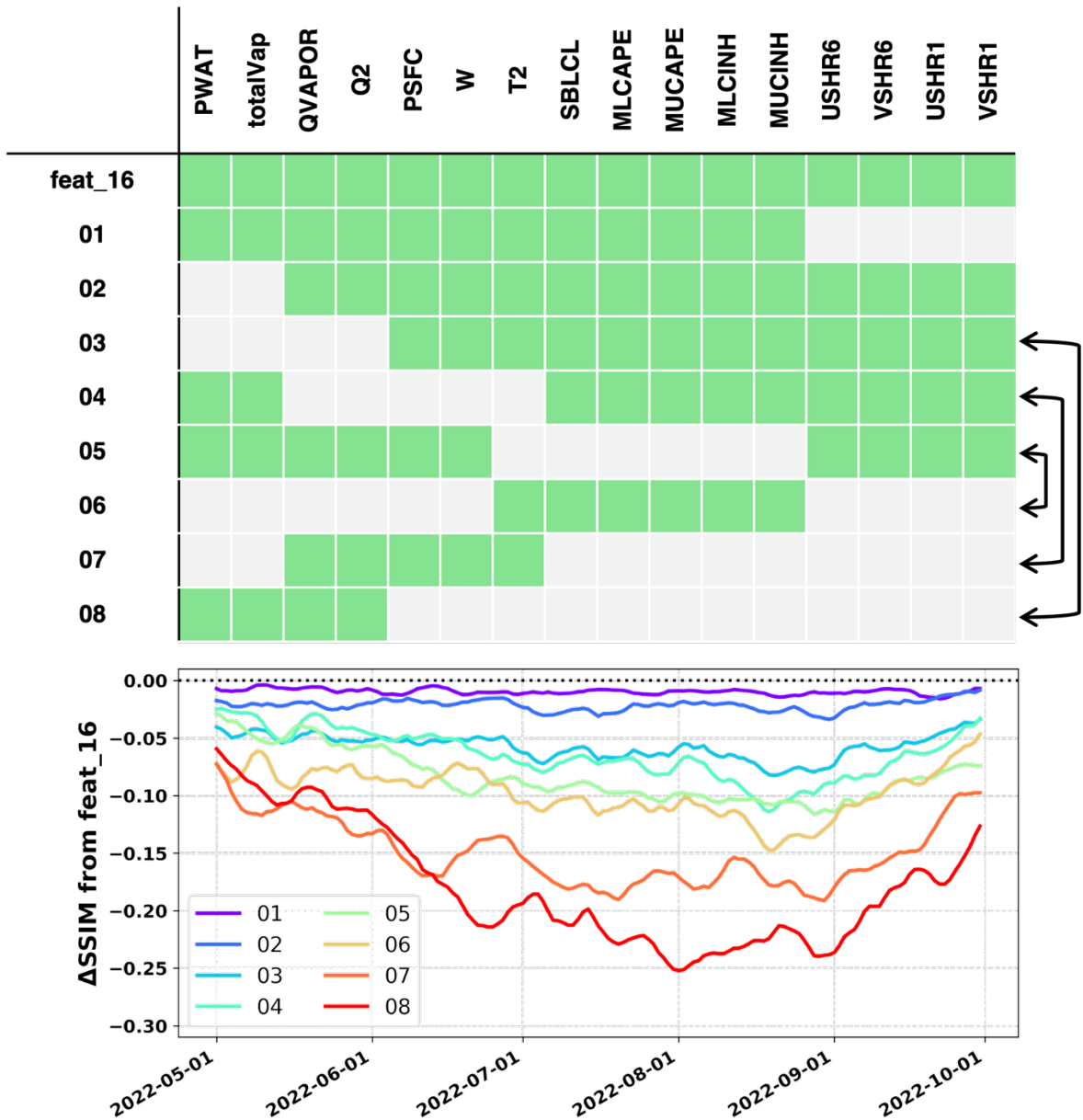

Figure 13. Top panel: Model configurations used in this ablation study. Each green box indicates the environmental variables included in the training, labeled by their corresponding ID on the left. Double arrow lines indicate paired configurations with opposite sets of input features. Bottom panel: SSIM differences between each ablated model and the full model trained with all 16 environmental variables (feat_16), evaluated during the warm season of the testing period (May–September 2022).

## 5. Conclusions

In this study we demonstrate that machine learning (ML) techniques such as *multilayer perceptron (MLP) models can effectively predict sub-grid-scale convective cloud information, such as convective frequency, from environmental variables alone.* MLP models trained on coarsened 28-km grid data exhibit robust performance across different initializations and preprocessing strategies, consistently outperforming baseline linear regression models. The model reliably captures spatial structures, regional hotspots, and diurnal variability of convective activity, even without explicit spatial or temporal information. Importantly, the MLP models generalize well to unseen years and climate regimes, including pseudo-global-warming (PGW) scenarios.

These findings highlight the potential for pixel-based MLP approaches to bridge the scale gap between coarse-resolution climate model outputs and sub-grid convective processes typically unresolved in general climate models (GCMs), such as the well-known challenges GCMs face in representing the convective diurnal cycle. The demonstrated ability of MLP models to retrieve realistic diurnal convection cycles from coarse environmental variables underscores their utility in advancing climate diagnostics.

To better understand how MLP models relate sub-grid-scale cloud information to environmental variables, we conducted targeted experiments using regional generalizability and feature ablation tests. Regional generalizability experiments reveal that MLP models trained on environmental data from specific regions (e.g., north vs. south, east vs. west) exhibit limited skill when applied to distinct regions characterized by different convective regimes or surface conditions. Notably, models trained on data from mountainous or subtropical environments struggle to generalize to mid-latitude continental regions, and vice versa. These results demonstrate the different localized nature of convective–environment interactions and mesoscale convective processes over different regions and emphasize the importance of geographically diverse environmental variables when developing broadly applicable models for sub-grid convection.

The ablation study systematically removed groups of related input variables to assess their contributions to the model performance. The results further reveal that thermodynamic indices (e.g., CAPE, CINH, and LCL) are among the most influential predictors for capturing convective spatial frequency, with moisture and near-surface variables also providing important information. However, no single variable group is sufficient alone; interactions among features are critical, highlighting the inherently interacting nature of the environmental–convection relationship. Wind shear, although less impactful in isolation, demonstrates value through its interactions with other environmental predictors.

This study demonstrates the promise of machine learning, particularly a simple, pixel-based MLP architectures, in predicting sub-grid convective cloud information from large-scale environmental variables. Our findings highlight the critical role of environmental nonlinearity and feature interactions in shaping convective processes and determining model skill. Moving forward, we will apply these methods to coarser-resolution reanalysis data or GCM outputs to retrieve convective proxies where such processes are typically underrepresented.

To further advance this work, we plan to leverage interpretable machine-learning frameworks to deepen our understanding of the environmental mechanisms driving regional and temporal variability in convection. For example, Explainable Boosting Machines (EBMs) have already provided valuable insights into feature importance across different fields including atmospheric science. Building on these advanced techniques, future efforts should focus on uncovering how interactions among environmental factors shape convective processes, with the broader goal of improving both the physical understanding and practical modeling of convection in present and future climate systems.

*Acknowledgments.*

This research was supported by the U.S. National Science Foundation (NSF) Collaborations in Artificial Intelligence and Geosciences (CAIG) program (grant number: 2425923). K. Rasmussen was also supported by National Oceanic and Atmospheric Administration (NOAA) grant (grant number: NA24OARX431C0055-T1-01). The authors acknowledge the use of the CONUS404 simulations, produced and made publicly available by the NSF National Center for

File generated with AMS Word template 2.0

Atmospheric Research (NSF NCAR). High-performance computing resources were provided by NSF NCAR's Computational and Information Systems Laboratory (CISL) Derecho: HPE Cray EX System (University Community Computing) through a small allocation project (project number: UCSU0090).

We thank Dr. Rick Schulte for valuable discussions and feedback throughout this work. We also acknowledge the open-source software community, whose tools, including Python, PyTorch, xarray, and matplotlib, made this research possible.

*Data Availability Statement.*

The primary dataset used in this study is the CONUS404 regional climate simulation, produced by NSF NCAR. The CONUS404 simulations are publicly available through the NSF NCAR Research Data Archive at https://doi.org/10.5065/ZYY0-Y036.

The algorithms and codes for convective-stratiform masking and cloud systems classification have been archived on Zenodo (https://doi.org/10.5281/zenodo.6491940). Processed datasets, including the coarsened environmental variables and convective frequency fields used for machine learning training and evaluation, are available upon request from the corresponding author. All machine learning training, evaluation, and preprocessing codes are currently archived in a private GitHub repository and available upon request. The code will be made publicly accessible once the paper is published.

# APPENDIX

## Appendix Title

NA

File generated with AMS Word template 2.0